\documentstyle[aps,prl,epsfig,multicol]{revtex}
\epsfclipon

\begin{document}

\newcommand \be {\begin{equation}}
\newcommand \ee {\end{equation}}
\newcommand \bea {\begin{eqnarray}}
\newcommand \eea {\end{eqnarray}}
\newcommand \nn {\nonumber}

\title{\bf The Kovacs effect in model glasses}
\author{E.M. Bertin$^1$, J.-P. Bouchaud$^1$, J.-M. Drouffe$^2$, C. Godr\`eche$^{1,3}$}
\address{\it
$^1$ Service de Physique de l'\'Etat Condens\'e, CEA Saclay, F-91191 Gif-sur-Yvette Cedex, France\\
$^2$ Service de Physique Th\'eorique, CEA Saclay, F-91191 Gif-sur-Yvette Cedex, France\\
$^3$ Dipartimento di Fisica, Universit\`a di Roma ``La Sapienza'' and
Center for Statistical Mechanics and Complexity,\\
INFM Roma 1, Piazzale AldoMoro 2, I-00185 Roma, Italy
}

\maketitle

\begin{abstract}
We discuss the `memory effect' discovered in the $60$'s by Kovacs in temperature shift
experiments on glassy polymers, where the volume (or energy) displays a non 
monotonous time behaviour.  This effect is generic and is observed on a variety of 
different glassy systems (including granular materials). The aim of this paper is to 
discuss whether some microscopic information can be extracted from a quantitative 
analysis of the `Kovacs hump'. We study analytically two families of theoretical models: domain growth and traps, 
for which detailed predictions of the shape of the hump can be obtained. 
Qualitatively, the Kovacs effect reflects the heterogeneity of the system: its description 
requires to deal not only with averages but with a full 
probability distribution (of domain sizes or of relaxation times). We end by some suggestions for a quantitative analysis of experimental results.
\end{abstract}

\section{Introduction. The Kovacs effect}

Systems with slow or glassy dynamics often exhibit non trivial behaviour 
when temperature changes are applied within the glassy phase. Since the system
is out of equilibrium, one expects that its properties generically depend on 
the history of the system, an effect that is often called `memory'. However, this
general term embraces rather different effects. In the recent spin-glass literature, 
memory is associated to a {\it two time} observable, such as the a.c. susceptibility
(that depends both on the frequency and on the age of the system) or any other 
response function. It has been shown
that after a negative temperature cycle, the a.c. susceptibility recovers the exact
value it had before the negative temperature jump, hence the name memory. This
effect would be trivial if the dynamics was totally frozen at low temperature, whereas
experiments show very clearly that some noticeable evolution in fact takes place 
\cite{memory1,memory2,memory3}. The same qualitative effect, although not as clear-cut
as in spin-glasses, has been observed in many other glassy materials (polymers, colloids,
ferro-electrics, etc.) \cite{PMMA,levelut,Bohmer,gelatine,munch}. 

There is however another well known `memory effect' that was discovered by Kovacs fourty years ago. This effect concerns {\it one time} observables, such as the 
specific volume, or the energy density, etc. and clearly shows that the non equilibrium 
state of the system cannot be fully characterized by the (time dependent) value 
of thermodynamical variables. The procedure followed by Kovacs was the 
following \cite{Kovacs63}: first, a reference curve is obtained by quenching the sample 
from a high temperature $T_0$ to a low temperature $T_2$, and measuring the 
time dependent volume $V(t)$ until a 
time $t_{eq}$ where the system can be considered to be in equilibrium. This defines a volume 
$V_{eq}(T_2) = V(t_{eq})$. In a second step, the sample is quenched again from $T_0$ to 
a temperature $T_1 < T_2$, until a certain time $t_1$. The 
temperature is then quickly raised from $T_1$ to $T_2$. The time $t_1$ is chosen
such that the volume just after the jump reaches the value 
$V(t_1^+) = V_{eq}(T_2)$ -- whereas in equilibrium ($t_1 \to \infty$) one would have $V_{eq}(T_1) < V_{eq}(T_2)$. 
Naively, one 
expects that nothing should happen, since the volume is already at its `correct' equilibrium 
value $V_{eq}(T_2)$ at the new temperature. The volume $V(t)$ 
in fact shows a {\it non monotonic behaviour} for $t>t_1$, first 
increasing and then relaxing back to the equilibrium value $V_{eq}(T_2)$:
\be
V(t) = V_{eq}(T_2) + \Delta V(t),
\ee 
where $\Delta V \geq 0$ is the `Kovacs hump', such that $\Delta V(t=t_1^+)=0$ and 
$\Delta V(t \to \infty)=0$. Note that the condition $V(t_1^+) = V_{eq}(T_2)$ (and not $V(t_1^-) = V_{eq}(T_2)$) is chosen as to
remove the trivial part of the effect, due to the thermal expansion of the fast (local) degrees of freedom. 
This subtlety in the Kovacs protocol is in fact quite important, as will be clear below.  

\begin{figure}
\centerline{
\epsfxsize = 8.5cm
\epsfbox{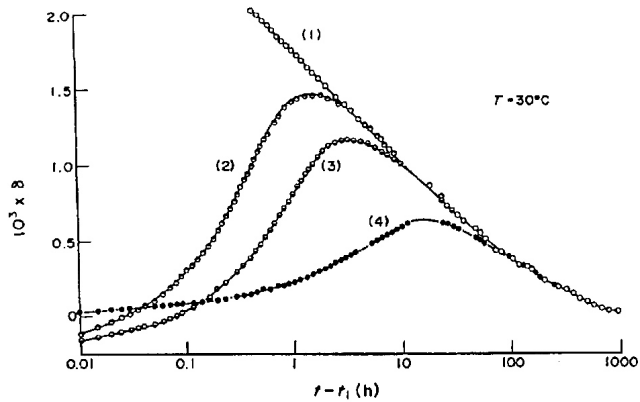}
}
\vskip 0.5 cm
\caption{\sl Isothermal evolution at $T_2=30^{\circ}C$ of the relative variation of the volume ($\times 10^3$) in polyvinyl acetate: after a direct quench from $T_0=40^{\circ}C$ to $T_2=30^{\circ}C$ (1); after quenches from $T_0=40^{\circ }C$ to $T_1=10^{\circ}C$ (2), $15^{\circ}C$ (3), or $25^{\circ }C$ (4) followed by rapid re-heating at $T_2=30^{\circ}C$. Data taken from A.~J.~Kovacs, Adv. Polym. Sci. {\bf 3}, 394 (1963).}
\label{figkovacs}
\end{figure}

Fig.~\ref{figkovacs} reproduces the 
original results published by Kovacs in 1963 \cite{Kovacs63}, obtained on polyvinyl acetate.
The Kovacs effect shows unambiguously that other `internal' variables, beside the volume,
are needed to characterize the out of equilibrium state of the system, and that these variables did not reach their 
$T_2$ equilibrium value at the end of the first stage. The memory in this case refers 
to the fact that these internal variables keep track, to some extent, of the system
history. To avoid confusion between the different types of memory effects, we will follow \cite{Berthier} and call the above phenomenon the Kovacs effect. 
The Kovacs hump is characterized by its height $\Delta V_K$, and by the time $\tau_K$ 
for which the maximum is reached: $\Delta V(t=t_1+\tau_K)=\Delta V_K$. Qualitatively, 
the height $\Delta V_K$ {\it grows} with the temperature difference $T_2 - T_1$ (it
should obviously be zero for $T_1=T_2$), whereas the time $\tau_K$ {\it decreases} when 
$T_2 - T_1$ increases.

A similar effect was recently reported in the context of granular materials \cite{Josserand}.
In the first stage of another type of experiment one `taps' the system with three different 
amplitudes --say weak, moderate and 
strong-- during a time chosen such as to reach a certain density, identical in the
three cases. In the
second stage of the experiment, the tapping amplitude is chosen to be moderate. 
The density just after the amplitude `jump' is recorded. If the state of the system was 
only described by its density, the evolution of the density after the jump should be identical 
for all three situations, and follow the `moderate' reference curve. This is not the case:
as for the polymer glass, the weakly tapped system first has to {\it dilate} before it 
is able to resume its compaction, whereas the strongly tapped system compacts faster than the 
reference system just after the jump \cite{Josserand}.

Finally, the same effect was recently observed in a numerical simulation of three dimensional 
spin-glasses \cite{BB02} and in a realistic model of molecular liquid \cite{Sciortino}.
In spin-glasses, the energy density reveals the characteristic Kovacs hump when the 
temperature is raised ; the height of the hump and the time of the maximum 
behave qualitatively as in polymer glasses. Features similar to the Kovacs effect have 
also been identified experimentally in dipolar glasses \cite{Alberici} and spin glasses 
\cite{munetakaetal}. Since the Kovacs effect seems to be rather ubiquitous, a natural question is whether the
underlying physics is the same in all these systems. Stated differently, can the effect
{\it select} between different microscopic models of glassy dynamics?

The aim of this somewhat didactic paper is to discuss some simple models that allow
to shed light on the above questions. In these models, the `internal' variables referred to
above appear as a whole distribution function (of domain sizes, or of relaxation times) of
which only the {\it mean} is fixed by the experimental protocol, whereas the {\it shape} of the
distribution keeps track of the system history. We show that the Kovacs effect is indeed rather
generic, but that the detailed shape of the `Kovacs hump' could reveal some useful 
microscopic information on the underlying glassy dynamics (see also the 
discussion in \cite{BerthierHouches}). 
We first discuss models where slow dynamics
is due to a coarsening mechanism, and recall and generalize the main results of \cite{Berthier}.
We then turn to the Kovacs effect in the trap model, where  detailed calculations
can be performed. We end the paper with some suggestions for
further analyzing experimental results, with the hope that the Kovacs effect could help
identifying distributions of relaxation times, and/or
provide some indirect evidence for a growing length scale in glassy systems.  

\section{The Kovacs effect and domain growth}

\begin{figure}
\centerline{
\epsfxsize = 7.5cm
\epsfbox{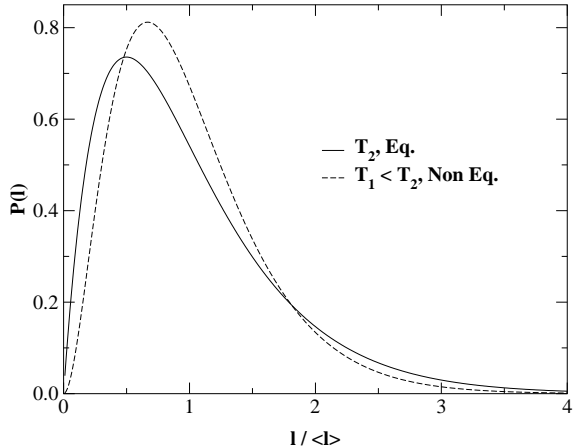}
}
\vskip 0.5 cm
\caption{\sl Distribution of domain sizes in the one-dimensional Ising model corresponding to two different temperatures, 
such that the
average domain size is identical in the two cases (from \protect\cite{Brawer}). The out of equilibrium distribution 
$T_1 < T_2$ (dotted line) 
is more sharply peaked than the corresponding equilibrium distribution at $T_2$
(plain line). Upon heating, small domains, initially less numerous, quickly appear within large ones.}
\label{figPell}
\end{figure}

The simplest out of equilibrium system is the one-dimensional Ising model with Glauber
dynamics. This system does not order at any non zero temperature, but at sufficiently
low temperatures the equilibrium domain size $\xi$ becomes large and for times shorter than the equilibration time, 
the dynamics is governed by the
growth of the typical domain size as the square root of time.
The energy, which is simply related to the average density of domain walls, plays in this model the r\^ole of the volume in Kovacs' experiments. When the system is prepared at $T_1$ 
for a time $t_1$ such that the average distance between the walls is equal to the
equilibrium size at $T_2 > T_1$, the out of equilibrium distribution of domain sizes at $T_1$ 
is more sharply peaked around its mean than the corresponding equilibrium distribution at $T_2$ --see Fig.~\ref{figPell}. 
In particular, the number of small domains is depleted from its equilibrium value. Upon heating, the first effect is that 
some extra domain walls 
nucleate within the larger domains, causing the number of small domains (and the energy) to increase. The 
exact shape of Kovacs' hump can be computed in this model \cite{Brawer}, and is found to be {\it linear} 
in time for small times, with a slope that increases with the temperature difference $T_2-T_1$,
before reaching the (exponential) relaxation curve describing a simple quench from high 
temperatures. Note that the relaxation time is finite for all $T > 0$ in this model; 
the rate of the final decay only depends on $T_2$, but not on $T_1$. As discussed by 
Brawer \cite{Brawer}, this is qualitatively similar to the experimental curves reported in Fig.~\ref{figkovacs}.

In systems where the equilibrium domain size is infinite, or very large compared to 
the dynamical length corresponding to the experimental time scale,\footnote{If the equilibrium 
correlation length is small, then the Kovacs effect is trivial in the sense that only fast 
degrees of freedom need to reequilibrate. Only if the maximum of the Kovacs hump occurs at times 
much larger than the atomic time scales will one observe a non trivial effect when following the 
Kovacs protocol, where the contribution of fast degrees of freedom is removed by choosing $E(t_1^+) = E_{eq}(T_2)$. 
(See the discussion in the introduction).} the mechanism for the 
Kovacs effect in coarsening systems is more generally the following \cite{Berthier,BB02}:
after a time $t_1$ spent at $T_1$, the system orders up to a scale $\ell_1=\ell(t_1,T_1)$, 
leading
to an excess energy (over the bulk contribution) due to the presence of domain walls with typical scale $\ell_1$.
This excess energy density behaves as $\ell_1^{\Theta-d}$, where $\Theta$ is the exponent giving the scaling 
of the excess  energy of a domain with its size (for example, $\Theta=d-1$ for the Ising model, and $\Theta=d-2$ 
in the XY-model).
When the temperature is increased to $T_2$, the bulk energy density (within the domains) is 
suddenly too low compared to the equilibrium value at $T_2$. This bulk contribution to the 
energy density therefore increases rapidly by nucleating new domain walls within 
the large preexisting domains of size $\ell_1$. This picture was actually suggested in \cite{Alberici}
to interpret an `overshoot' effect in dipolar glasses which, with 
hindsight, is the
precise counterpart of the Kovacs effect in these materials.

For larger times,
the primary coarsening process resumes and the density of domain walls
decreases, leading to a decrease of the total energy density. This decay
is a priori expected to dominate when the length scale $\ell(t,T_2)$ associated to the 
dynamical processes initiated by the temperature change becomes of the order of
$\ell_1$, i.e. after a time $\tau^*$ such that:
\be
\ell(\tau^*,T_2)=\ell_1.
\ee
However, the time $\tau_K$ at which the maximum of the Kovacs hump occurs 
turns out to be, in general,
much smaller than $\tau^*$ (but still much larger than the microscopic time scale $\tau_0$). 
More precisely, the following picture emerges
from the exact computation of \cite{Berthier}:

\begin{itemize}

\item In the limit where $\ell(t,T_2)$ and $\ell_1$ are much larger than the lattice
spacing $a$, the fast initial nucleation processes have taken place, 
and one can expect the
energy density $E$ to take the following scaling form:
\be\label{dg}
\Delta E = E(t_1+t) - E_{eq}(T_2) = \Delta E_K \, {\cal 
F}\left(\frac{\ell(t,T_2)}{\ell_1}\right),
\qquad \ell(t,T_2) \gg a
\ee
where $\Delta E_K$ is the height of the Kovacs' hump, and  ${\cal F}(u) \sim 
u^{\Theta-d}$ when $u \to \infty$. Using the fact that $\Delta E$ should not depend on 
$t_1$ at large times, one finds $\Delta E_K \sim \ell_1^{\Theta-d}$, which means that
the energy scale of the hump is of the order of the excess energy stored in the domain walls
at $T_1$. As shown in \cite{Berthier}, the above scaling form indeed 
holds exactly
for the 2D XY model in the ordered critical phase, for which $\Theta=d-2$. One 
finds in that case ${\cal F}_{XY}(u) = (1+u^2)^{-1}$.

\item In the short time limit $\ell(t,T_2) \sim a$, one expects a nucleation contribution to 
$\Delta E$ responsible for the Kovacs hump. In the case of the critical XY model, where the 
thermal correlation length $\xi$ is infinite, one finds a power-law contribution \cite{Berthier}:
\be
\Delta E \approx \Delta E_K \left[1 -\left(\frac{\ell(t,T_2)}{a}\right)^{\Theta-d}\right].
\ee
Note that this contribution vanishes for $t=0$, since $\ell(t=0,T_2)=a$, but 
{\it cannot} be written as a scaling function of $\ell/\ell_1$. This is at the 
origin of the difference between $\tau_K$ and $\tau^*$. The above results only hold if $\ell_1 \ll \xi$. 
In the other limit where the correlation length $\xi$ is small,
the above power-law is replaced by a fast exponential convergence. In this case $\tau_K \sim \tau_0$,
and the Kovacs effect becomes trivial (it would actually disappear if the Kovacs protocol was used -- see the previous
footnote).

\end{itemize}

In this domain growth scenario, one finds the length scale 
$\ell_1$ (and therefore $\tau^*$) to be a decreasing function of $T_2 - 
T_1$. Physically, this is
because the bulk energy contribution is lower for smaller $T_1$; the residual domain 
wall energy density ($\sim \ell_1^{\Theta-d}$) must then be larger
in order to ensure that in the Kovacs protocol, the time $t_1$ is determined such that:
\be
E(t_1,T_1)=E_{eq}(T_1)+ \ell_1^{\Theta-d} = E_{eq}(T_2).
\ee
For small $T_2 - T_1$, one thus expects a linear relation $\Delta E_K \sim \ell_1^{\Theta-d} \propto C (T_2 - T_1)$, where
$C$ is the specific heat. Therefore, the qualitative dependence of both $\tau^*$ and $\Delta E_K \sim \ell_1^{\Theta-d}$ with $T_2 - T_1$ is correctly predicted by this picture.

If the length $\ell(t,T)$ grows as a power of time, then from Eq.~(\ref{dg}), $\Delta E/\Delta E_K$ is found to be a scaling function of $t/\tau^*$ in the limit of large times, where the initial (non scaling)
contribution due to nucleation vanishes. Due to this non-scaling contribution, the 
scaling function ${\cal F}$ has a non zero value for small arguments: ${\cal F}(0^+) > 0$. Therefore, in the 
domain growth scenario (including the equilibrium case
discussed by Brawer and recalled above), the Kovacs hump does {\it not} rescale as a 
function of $t/\tau_K$, because the position of the maximum $\tau_K$ is determined 
by the non scaling nucleation contribution.
The time should rather be rescaled by $\tau^*$ determined such that the amplitude of the hump has decreased by a factor two
(say). By the same token, one expects to see systematic deviations from scaling in the regime $t \ll \tau^*$, due to the non scaling contribution of nucleation 
process.

We now turn to another soluble model that, interestingly, predicts a 
variety
of shapes for the Kovacs hump, which in some regimes are very similar 
to the ones predicted by the domain growth model.

\section{The Kovacs effect in the trap model}

\subsection{Definition of the model}

A simple model exhibiting glassy behaviour is the trap model, 
which has been extensively studied in the literature \cite{Bouch92,Dean95,Monthus},
and generalized to describe the rheology of soft glassy materials \cite{Sollich}, or the dynamics of contacts in granular media \cite{Kabla}.
In this model, a particle is trapped in potential wells, and can escape only 
through thermal activation. The depth (energy barrier) of the well is a random variable $E > 0$ with an exponential a priori distribution $\rho(E) = T_g^{-1}\, e^{-E/T_g}$. 
When the particle is in a trap $j$ of energy $E_j$, it will escape after a time $\Delta t$ 
distributed according to $p_j(\Delta t) = \tau_j^{-1}\, e^{-\Delta t/\tau_j}$, 
where $\tau_j = \tau_0 \, e^{E_j/T}$ is the mean trapping time of the site $j$, and 
then chooses a new trap among all the others with a uniform probability. The microscopic time scale $\tau_0$ is taken as the time unit in the following. 
The energy scale $T_g$ turns out to be also the phase transition temperature. 
For $T > T_g$, the system equilibrates and behaves like a `liquid', 
whereas for $T < T_g$, the lowest energy states become the most probable ones 
and the system never stops aging. Of course, this model should not be considered 
as a realistic microscopic model, but rather as a coarse-grained phase-space model --see the discussion in \cite{Houches,BenArous}.
Also, an exponential distribution of energies might not be the most appropriate 
description of a given system. For example, recent simulations of Lennard-Jones systems
\cite{Reichman} have shown that a Gaussian distribution of barriers is in fact more adequate.
As noted in \cite{Monthus}, the results of the exponential trap model can be extended to that
case.

Due to its simplicity, this model allows one to obtain analytic expressions of many quantities of 
interest. As for coarsening models, we have chosen the energy as the natural observable
that plays the r\^ole of the volume in Kovacs' experiments.

Let us now present the explicit calculation of the energy as a function of time, 
with the temperature protocol defined in the introduction. However, since fast degrees of freedom are 
absent in the trap model (there is no `bottom of the wells' dynamics), one does not need to
distinguish between $t_1^-$ and $t_1^+$, as is important both experimentally and in models with 
microscopic degrees of freedom (see above for a discussion of this point). Two different 
cases have been considered in details. In the first one, the temperatures $T_1$ and $T_2$ 
are both {\it above} $T_g$, but close to it, so that the system eventually equilibrates, 
but with very long relaxation times. In the second case, both temperatures are below 
$T_g$, so that the system is in the aging regime where equilibration is never achieved. 
Finally, we only briefly discuss the `mixed' case where $T_1<T_g<T_2$. The original Kovacs 
experiment corresponds to the first case, since the volume is seen to  
relax towards its equilibrium value at $T_2$, used as the reference energy.
In the second case, the time $t_1$ at which temperature is shifted is in fact arbitrary, 
but interesting scaling properties appear.

\subsection{Case $T>T_g$: relaxation towards equilibrium}

We shall use a continuous energy description (see \cite{Monthus}), i.e. the system is 
described by the probability $P_{\textsc{t}}(E,t)$ to be in a state with an energy (barrier) $E$ at time $t$ and temperature $T$, which 
evolves according to the following Master equation:

\be \label{Master-eq}
\frac{\partial P_{\textsc{t}}}{\partial t}(E,t) = -\, e^{-E/T} P_{\textsc{t}}(E,t) + \omega(t) \rho(E)
\ee
with $\omega(t) = \int_0^{\infty} dE' \, e^{-E'/T} P_{\textsc{t}}(E',t)$ is the 
average hopping rate.
For $T>T_g$, $P_{\textsc{t}}(E,t)$ relaxes towards the equilibrium distribution 
$P_{\textsc{t}}^{eq}(E) = Z^{-1} e^{E/T}$. So the interesting quantity to study 
is the deviation from equilibrium, i.e. the distribution $p_{\textsc{t}}(E,t)$ 
defined as $p_{\textsc{t}}(E,t) = P_{\textsc{t}}(E,t)-P_{\textsc{t}}^{eq}(E)$. Let us first focus on a 
simple isothermal quench from a given initial condition $P_0(E)=P_{\textsc{t}_0}^{eq}(E)$. 
The evolution of $p_{\textsc{t}}(E,t)$ 
can be computed using a time Laplace transform, and if $T_0 > T$, the asymptotic behaviour 
of the distribution 
becomes independent of the initial condition $P_0(E)$, yielding:
\be
\hat{p}(E,s) = \frac{(\beta_g-\beta)\, e^{-(\beta_g-\beta)E}}{1+s\,e^{\beta E}} [ \Gamma(\theta)\, \Gamma(2-\theta)\, s^{\theta-2} - e^{\beta E}]
\ee
where $\beta=1/T$ and $\theta=T/T_g$ is the reduced temperature. 
Let us define the energy deviation $\varepsilon_{\textsc{t}}(t)=|E_{\textsc{t}}(t)-E_{\textsc{t}}^{eq}|$. 
Note that in the following, energies are understood to be {\it true} physical energies, 
i.e. the opposite of the energy barriers: 
$\varepsilon_{\textsc{t}}(t) = -\int_0^{\infty} dE\, E\,  p_{\textsc{t}}(E,t)$. 
This last quantity can be computed from $\hat{p}(E,s)$, which gives:
\be \label{eqn-isoquench}
\varepsilon_{\textsc{t}}(t) = \frac{T}{t^{\theta-1}} [\Gamma(\theta)\, \ln t - \Gamma'(\theta)].
\ee
Hence, the energy relaxation above $T_g$ is (up to a logarithmic correction) a power law
with an exponent that becomes small for $T \to T_g$. The time $t_1$ when the temperature 
has to be raised from $T_1$ to $T_2$ in the Kovacs procedure is defined by 
$E_{\textsc{t}_1}(t_1) = E_{\textsc{t}_2}^{eq}$, or equivalently 
$\varepsilon_{\textsc{t}_1}(t_0) = E_{\textsc{t}_2}^{eq} - E_{\textsc{t}_1}^{eq}$. Thus $t_1$ is determined 
by the equation:
\be
\frac{1}{t_1^{\theta_1-1}} [\Gamma(\theta_1)\, \ln t_1 - \Gamma'(\theta_1)] \approx
\frac{\theta_2-\theta_1}{(\theta_1-1)^2}.
\ee
Note that in order to be consistent, the above equation assumes that 
$\theta_2-\theta_1 \ll (\theta_1-1)^2 \ll 1$, in which case $t_1 \gg \tau_0\, (=1)$.

Now using the distribution $p_{\textsc{t}_2}(E,t_0) = p_{\textsc{t}_1}(E,t_0) + P_{\textsc{t}_1}^{eq}(E) 
- P_{\textsc{t}_2}^{eq}(E)$ as initial condition in the Master equation, one can compute the 
further evolution of the energy at $T_2$ at time $t_1+t$. A time scale $\tau^*=t_1^{\gamma}$ 
naturally appears (with $\gamma=\theta_1/\theta_2$). Defining the energy variation 
$\Delta E(t) = E(t_1+t)-E(t_1)$, one finds in the short-time regime $1 \ll t \ll \tau^*$:
\be
\Delta E(t) \approx \frac{T_1}{t_1^{\theta_1-1}} \left[\ln t_1 + 
\frac{1}{\theta_1-1}\right] - \left(1+\frac{1}{t_1^{\theta_1-1}} \right) 
\frac{T_2}{t^{\theta_2-1/\gamma}} \left[\ln t + \gamma_E  \right] + 
\frac{T_2}{t^{\theta_2-1}} [\ln t + \gamma_E]
\ee
where $\gamma_E = -\Gamma'(1)$ is the Euler constant. Interestingly, this behaviour is very 
similar to that found for the coarsening process. Indeed, in the limit 
$\theta_2-\theta_1 \ll (\theta_1-1)^2$, the two power laws in the previous equation, 
$\theta_2-1/\gamma$ and $\theta_2-1$ are very close to each other, and the expression can be 
simplified as: 
\be\label{early}
\Delta E(t) \approx \frac{T_1}{t_1^{\theta_1-1}} \left([\ln t_1 + 
\frac{1}{\theta_1-1}] - \frac{1}{t^{\theta_2-1}} [\ln t + \gamma_E]\right).
\ee
Therefore, the maximum of the Kovacs hump is given (in the considered limit) 
by:
\be
\Delta E_K \approx \frac{T_1}{t_1^{\theta_1-1}} \ln t_1 \approx \frac{\theta_2-\theta_1}{(\theta_1-1)^2}.
\ee
The approach to this maximum is described by a power law of time with a
logarithmic correction. This is not very different from the coarsening model discussed in the previous section. Note that the height of the hump is again linear in $T_2 - T_1$ for small 
temperature differences, as was the case for domain growth. Note also that 
$\tau^*=t_1^{\gamma}$ is a decreasing function of $T_2-T_1$, in agreement with experimental 
results.

In the long time regime $t \gg \tau^*$, one recovers as expected the isothermal 
quench result Eq.~(\ref{eqn-isoquench}) at temperature $T_2$:

\be \label{late}
\Delta E(t) =  \frac{T_2}{t^{\theta_2-1}} [\Gamma(\theta_2)\, \ln t - \Gamma'(\theta_2)]
\ee
This late time result can again be put in a scaling form (up to logarithmic corrections):

\be
\Delta E(t) =  \Delta E_K \, {\cal G}\left(\frac{t}{\tau^*}\right)\qquad \tau^*=t_1^\gamma
\ee
but the early time regime Eq.~(\ref{early}) fails to scale. Only when $T_1,T_2 \to T_g$, 
does one find that the maximum time $\tau_K$ coincides with $\tau^*$. More generally, and as
for domain growth, one has $\tau_K \ll \tau^*$. 

\subsection{The aging case ($T_1,\,T_2<T_g$)}

We now turn to the aging case where the shape of the hump is found to be 
qualitatively different. We consider the case where both $T_1$ and $T_2$ are 
less than $T_g$. In this case, the system never converges to an equilibrium state, 
but keeps on aging, so that the situation is different from that of the 
Kovacs original experiment, but could in principle also be investigated experimentally.
Since the equilibrium energy at $T_2$ does not exist, we choose to shift the temperature
from $T_1$ (initially reached at $t=0$) to $T_2$ after a waiting time $t_w$ (which plays the
role of $t_1$ in the previous sections).

\subsubsection{Probability distribution and Green function}

The continuous energy Master equation~(\ref{Master-eq}) does not admit anymore a stationary 
solution. The resulting dynamical distribution can be computed using the Laplace
transform $\hat{P}_{\textsc{t}}(E,s_w)$ (in the time domain) of $P_{\textsc{t}}(E,t_w)$, where $t_w$ 
is the waiting time since the quench from high temperatures. One finds, 
in the asymptotic regime $s_w \to 0$ (or $t_w \to \infty$):

\be
\hat{P}_{\textsc{t}}(E,s_w) \simeq \hat{\Pi}_{\textsc{t}}(E,s) \equiv \frac{\sin \pi \theta}{\pi}\, 
\frac{\beta\, e^{\beta E}}{(1+s\, e^{\beta E})(s\, e^{\beta E})^{\theta}},
\ee
where $\theta \equiv T/T_g$. Since $s\,\hat{\Pi}_{\textsc{t}}(E,s)$ is a function of the product $s\, e^{\beta E}$, 
$\Pi_{\textsc{t}}(E,t)$ depends only on the scaling variable $\xi = e^{\beta E}/t$.
Then one can turn to the computation of the energy variation after the temperature shift.

\begin{figure}
\centerline{
\epsfysize = 6cm
\epsfbox{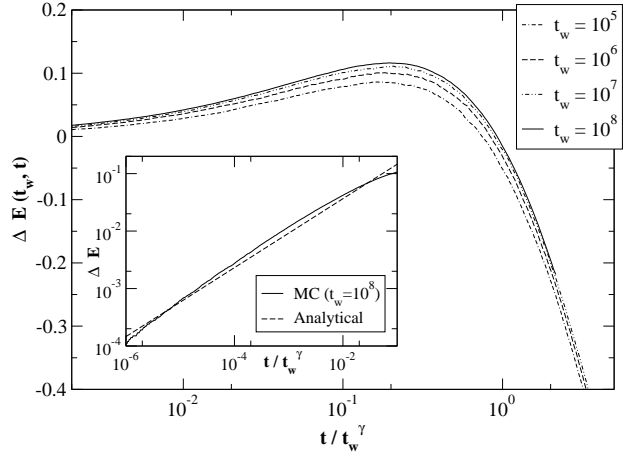}
}
\vskip 0.2 cm
\caption{\sl Plot of $\Delta E(t_w,t)$ in the trap model as a function of the scaling variable 
$t/t_w^{\gamma}$ for $t_w = 10^5$, $10^6$, $10^7$ and $10^8$ 
(Monte Carlo data), $\theta_1=0.5$ and $\theta_2=0.6$. Inset: 
comparison between Monte Carlo data ($t_w = 10^8$) and the analytical prediction
of the short time behaviour --see Eq.~(\ref{short-time1})-- for the same 
temperatures as above.}
\label{fig-trap5060}
\end{figure}

\subsubsection{Computation of the energy variation}

The detailed calculation of the energy variation $\Delta E(t_w,t)$ between time $t_w$ 
(when temperature is shifted from $T_1$ to $T_2$) and $t_w+t$ is given in Appendix A. 
Here we shall only summarize the main steps of the calculation, and emphasize physical 
interpretations and conclusions.
From a technical point of view, a useful tool in order to compute $\Delta E(t_w,t)$ is
the Green function $G_{\textsc{t}}(E,E_0,t)$ defined as the probability to have the energy 
$E$ at time $t_w+t$ given that the energy was $E_0$ at time $t_w$. This Green function
is computed, as for $P_{\textsc{t}}(E,t_w)$, using a Laplace transform, with $s$ the 
Laplace variable. One finds, for 
$s \tau_0 \ll 1$ ($t \gg \tau_0$), the following asymptotic expression:
\be
\hat{G}_{\textsc{t}}(E,E_0,s) = \frac{e^{\beta E_0}}{1+s\, e^{\beta E_0}}\, \delta(E-E_0) + 
\frac{1}{1+s\, e^{\beta E_0}}\, \hat{\Pi}_{\textsc{t}}(E,s)
\ee
Thanks to the Markovian properties of the dynamics, the Green function does not
depend on $t_w$, but only on the time difference $t$. One can express the average energy $\overline{E}(t_w+t)$ at time $t_w+t$ using the 
Green function:
\be
\overline{E}(t_w+t) = -\int_0^{\infty} dE \int_0^{\infty} dE_0\, E\, G_{\textsc{t}_1}(E,E_0,t) 
P_{\textsc{t}_2}(E_0,t_w)
\ee
where the minus sign accounts for the fact that the variable $E$ (i.e. the energy barrier) is actually the opposite of the true energy. The energy variation $\Delta E(t_w,t)$ is then:
\bea
\Delta E(t_w,t) &\equiv& \overline{E}(t_w+t) - \overline{E}(t_w)\\
	&=& -\int_0^{\infty} dE \int_0^{\infty} dE_0 (E-E_0) G_{\textsc{t}_1}(E,E_0,t) 
	P_{\textsc{t}_2}(E_0,t_w)
\eea
After a few calculations (see Appendix A), one can show that $\Delta E(t_w,t)$ exhibits a 
kind of `sub-aging' scaling (see also \cite{munetaka}):

\be\label{scaling-aging}
\Delta E(t_w,t) = \psi \left(\frac{t}{t_w^{\gamma}}\right),
\ee
where $\gamma=T_1/T_2 < 1$. One sees that the energy evolves on a 
typical time scale given by $\tau_K=\tau^*=t_w^{\gamma}$, which is expected from a simple activation 
argument. 
One can also study the asymptotic (short time and late time) behaviour of this scaling 
function. Concerning the short time behaviour, it is necessary to distinguish between two cases.
\begin{itemize}
\item If $\gamma>1-\theta_1$ (small temperature shifts) then $\Delta E(t_w,t)$ 
is found to be {\it singular} at short times:
\be \label{short-time1}
\Delta E(t_w,t) \simeq K_>\, \left( \frac{t}{t_w^{\gamma}} \right)^{(1-\theta_1)/\gamma} 
\qquad t \ll t_w^{\gamma}
\ee

\item If on the contrary $\gamma < 1-\theta_1$, one finds a linear $t$ dependence 
in the short time regime (with logarithmic corrections):
\be \label{short-time2}
\Delta E(t_w,t) \simeq K_< \left(\ln \frac{t_w^{\gamma}}{t}+C \right) 
\frac{t}{t_w^{\gamma}} \qquad t \ll t_w^{\gamma}
\ee
\end{itemize}
The coefficients $K_>$, $K_<$ and $C$ appearing in Eqs.~(\ref{short-time1},
\ref{short-time2}) are given in Appendix A --see Eqs.~(\ref{coef-K>}, \ref{coef-K<}, \ref{coef-C})-- and are found to be positive for $\theta_1 < \theta_2$. Therefore $\Delta E(t_w,t)$ is positive for short times, and the Kovacs effect has the expected sign.

\begin{figure}
\centerline{
\epsfysize = 6cm
\epsfbox{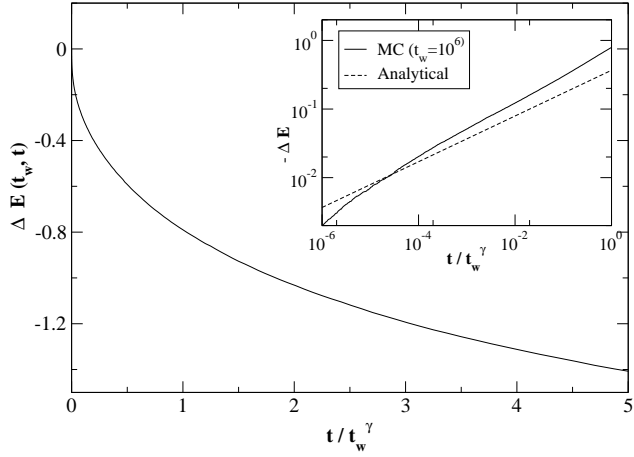}
}
\vskip 0.2 cm
\caption{\sl Plot of $\Delta E(t_w,t)$ in the trap model as a function of the scaling 
variable $t/t_w^{\gamma}$ for $t_w = 10^6$ (Monte Carlo data), in the 
case of a negative temperature shift: $\theta_1=0.6$ and $\theta_2=0.5$. 
Inset: comparison between Monte Carlo data and the analytical prediction 
of the short time behaviour --see Eq.~(\ref{short-time1}); note that finite size effects are strong in this case.}
\label{fig-trap6050}
\end{figure}

Note also that the coefficient $K_>$ vanishes linearly when
$\theta_1 \to \theta_2$. This is expected: if no temperature
jump occurs, the energy variation should be regular, i.e. 
linear in $t$. Moreover, if $\theta_1 > \theta_2$ 
(negative temperature shift, $\gamma >1$), the above
calculation is still valid, with a negative $K_>$,
and a non trivial exponent $(1-\theta_1)/\gamma$. The Kovacs hump becomes in this case 
a Kovacs trough.
Monte-Carlo data are compared with these analytical predictions in 
Figs.~\ref{fig-trap5060} and~\ref{fig-trap6050}, showing a rather good agreement. 
Note that the scaling Eq.~(\ref{scaling-aging}) is only approximate for finite $t_w$. 
A better rescaling can be obtained in the case $\theta_1 < \theta_2$ by plotting $\Delta E(t,t_w)/\Delta E_K$ as a function 
of $t/\tau_K$, where $\Delta E_K$ is the maximum value of $\Delta E(t_w,t)$, reached at $t=\tau_K \simeq t_w^\gamma$. Eq.~(\ref{scaling-aging}) means that asymptotically, $\Delta E_K$ becomes 
independent of $t_w$.

\begin{figure}
\centerline{
\epsfysize = 6cm
\epsfbox{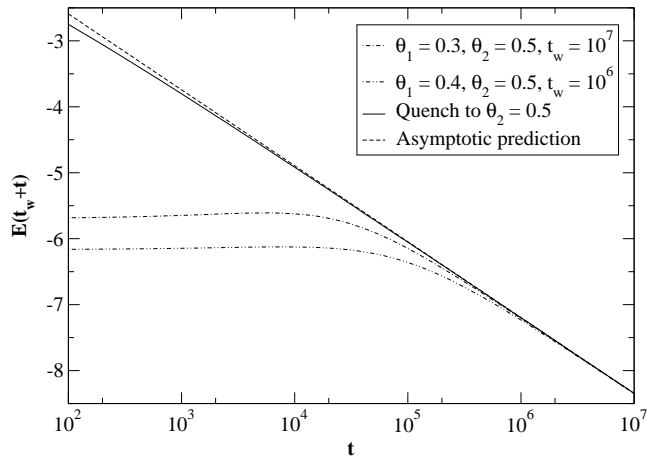}
}
\vskip 0.5 cm
\caption{\sl Comparison of the late time behaviour of $E(t_w+t) = \Delta E(t_w,t) + E(t_w)$ in the trap model
as a function of $t$, for temperatures $\theta_1=0.3$ and $0.4$, $\theta_2=0.5$
and waiting times $t_w=10^7$ and $10^6$ respectively (dot-dash) with a direct quench from 
infinite temperature to $\theta_2=0.5$ ($t_w=0$, full line). The asymptotic 
analytical prediction is also shown (dashed line).}
\label{fig-trap-late}
\end{figure}

Finally, the long time behaviour is easy to analyze: one can show that 
$P_{\textsc{t}}(E,t_w+t)$ behaves asymptotically in the same way 
whatever the initial condition $P_{\textsc{t}}(E,t_w)$. The
system behaves, at late time, as if it had been quenched directly from high 
temperature (see Fig.~\ref{fig-trap-late}). This means that in this limit
$\overline{E}(t_w+t)$ does not depend on $t_w$, but only on $t$:
\be
\overline{E}(t_w+t) \simeq \overline{E}_{late}(t) \equiv T_2\,
[\Gamma'(1) - \pi \cot \pi \theta_2] - T_2 \ln t
 \qquad t \gg t_w^{\gamma}
\ee
where $\overline{E}_{late}(t)$ is the average energy at a (large) 
time $t$ after a quench from a high temperature. So $\Delta E(t_w,t)$ 
is simply given by:
\be
\Delta E(t_w,t) = \overline{E}_{late}(t) - \overline{E}(t_w) \qquad t \gg t_w^{\gamma}
\ee

\subsection{The mixed case ($T_1 < T_g  <T_2$)}

For completeness, and since this is also an interesting situation, we briefly mention the results obtained in the trap model in the case where the glass transition temperature $T_g$ lies between $T_1$ and $T_2$: $T_1<T_g<T_2$. This case is worth studying, since the system can eventually equilibrate at the final temperature $T_2$, with long relaxation times (assuming $T_2$ is close to $T_g$), but $T_1$ can be varied in the whole range $0<T_1<T_g$ and not only in the vicinity of $T_g$. Interestingly, one finds the same short time singularities as in the aging case ($T_1<T_2<T_g$) studied above :
\bea
\Delta E (t_1,t) &\sim& \left( \frac{t}{t_1^{\gamma}} \right)^{(1-\theta_1)/\gamma} \qquad 1-\theta_1 < \gamma \\
\Delta E (t_1,t) &\sim& \frac{t}{t_1^{\gamma}} \qquad \qquad 1-\theta_1 > \gamma
\eea
with however prefactors and logarithmic corrections which are different from that found in the aging case.
In the long time regime, one naturally finds a convergence of $\Delta E (t,t_1)$ proportionnal to $t^{-(\theta_2-1)}$, as in the case $T_g<T_1<T_2$. As a result, one sees as could have been expected that the short time regime is generically dominated by the thermal history before the temperature shift, whereas the long time behaviour depends only on the final temperature $T_2$.

\subsection{Discussion -- `Fronts' in the energy distribution}

\begin{figure}
\centerline{
\epsfysize = 6cm
\epsfbox{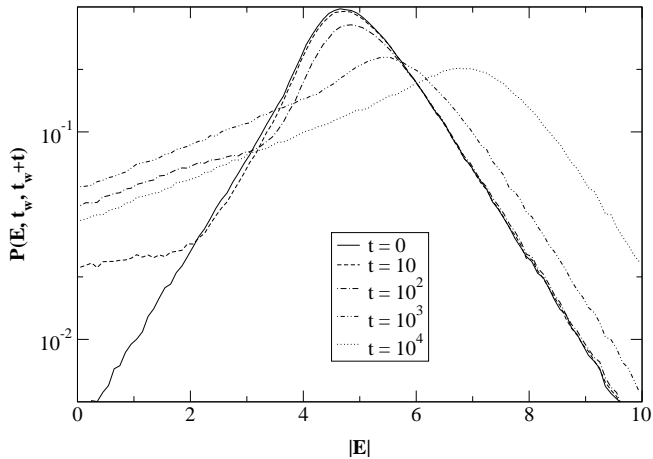}
}
\vskip 0.5 cm
\caption{\sl Dynamical energy distribution $P(E,t_w,t_w+t)$ at time $t_w+t$ after a (positive) temperature shift at $t_w$, with $t_w=10^4$, $\theta_1=0.5$ and $\theta_2=0.8$ (trap model). Time $t$ ranges from $t=0$ to $t=t_w$. One clearly sees the propagation of a `front' at small energies, associated to the re-equilibration of the short time scales, before the global drift of the distribution starts again. }
\label{distPE-fig}
\end{figure}

It is interesting to discuss how the distribution of energies $P(E,t)$ evolves 
when the temperature is shifted from $T_1$ to $T_2 > T_1$. This is illustrated in Fig.~\ref{distPE-fig}: at the lowest temperature, 
the probability of small (negative) energies is depressed. 
When the temperature is raised, the system obviously re-equilibrates fastest in the region 
of small energies since this corresponds to the smallest relaxation times.
The probability `hole' is thus rapidly filled, leading to an increase of the \
average energy, and thus to the Kovacs effect.  
As time increases, the equilibration progresses as a kind of `front' in energy 
space, as shown in Fig.~\ref{distPE-fig}. Only at later times does the peak of the distribution 
move to larger (negative) energies. It is interesting to realize that this picture is 
in fact very close the one emerging from the coarsening model where
short scales re-equilibrate fast and lead to an increase of the average energy, before
larger length scales resume the coarsening process (see Fig.~\ref{figPell}, and the 
discussion of Section II).  

The conclusion from the `domain growth' interpretation of the Kovacs effect 
presented in the previous section
is that a quantitative analysis of the Kovacs effect might give one a unique 
tool to investigate experimentally the problem of growing length scales 
or the statistics of trapping times in glassy systems, a topic of huge current interest 
\cite{donati98,donati98a,Weeks,Heuer,Reichman,Berthiernew,Biroli}. However, 
as demonstrated above, the quantitative predictions of the
trap model are in fact very similar to that of domain growth. As discussed recently in 
\cite{BerthierJPG}, the physical difference between the two pictures is 
not as obvious as it 
might first seem. In particular, the trap model description implicitly assumes the existence of an underlying 
`coherence length' \cite{Houches}; conversely, domain growth models may naturally generate 
a non trivial distribution of relaxation times \cite{BerthierJPG}. A possible
discrimination might lie in the temperature dependence of the short-time 
and long-time exponents that describe the Kovacs hump. While a temperature dependence is expected in an activated trap like description, it is less 
natural 
for power-law domain growth. On the other hand, more complicated (logarithmic) growth laws can
mimic power-laws with a temperature dependent exponent \cite{PRB,BB02}.

\section{Summary and Conclusions}

Although the Kovacs effect has been known for fourty years, its quantitative interpretation
has not been much developed until recently. In view of the fact that this effect is generic 
and observed in a variety of different `glassy' systems and models (such as the ones studied
in the present paper), it is important to 
establish which type of microscopic information one can extract from the quantitative 
analysis of the `Kovacs hump'. Qualitatively, the Kovacs effect reflects the {\it heterogeneity}
of the system: fixing the overall (macroscopic) value of the volume or energy does not
prevent the existence of local fluctuations, which keep track of the system history. A more
complete description of the system therefore requires to deal with  full distribution functions, 
and not only with averages. In the two models studied in this paper, this distribution 
function is that of {\it domain sizes} in domain growth models, and that of {\it relaxation 
times} in the trap model.
These models lead to precise, quantitative predictions for the shape of the Kovacs hump, 
which are summarized in Table~\ref{tab-recap}, and by the following phenomenological equation that describes the Kovacs hump [inspired from both domain growth and trap models --see Eqs.~(\ref{dg},\ref{early},\ref{late})]:
\be \label{fit}
\Delta E = \Delta E_K \left[\frac{\tau^{*\nu}}{(t+\tau^*)^\nu} - \frac{\varphi}{t^\nu}\right]
\ee 
In domain growth models, the exponent $\nu$ is equal to $(d-\Theta)/z$, where $z$ is the 
dynamical exponent relating length and time: $\ell \sim t^{1/z}$. The 
first term corresponds to the long time contribution of 
already grown domains, and the second term to the excess energy created by the transient 
nucleation of new domains. A similar interpretation of the two terms can be given 
within the trap model, with $\nu=T/T_g -1$ (for an exponential distribution of 
barriers).  Note that from Eq.~(\ref{fit}), one sees
that the data should re-scale as $\Delta E_K \, {\cal G}(\frac{t}{\tau^*})$ 
when $\tau^*$ 
becomes large. The position of the maximum $\tau_K$
extracted from Eq.~(\ref{fit}) is, in the limit $\tau^* \gg 1$, given by:
\be
\tau_K \approx (\varphi \tau^*)^{\frac{1}{\nu+1}} \ll \tau^*
\ee
so that the correct scaling variable is {\it not} $t/\tau_K$, except in the limit $\nu \to 0$, where
$\tau_K$ and $\tau^*$ coincide.

We hope that these results will motivate new, 
systematic experiments. It would in particular be valuable to test Eq.~(\ref{fit}). It would be very interesting 
to extract from a detailed analysis of the Kovacs effect a quantitative
determination of the distribution of relaxation times in glassy systems, and its temperature
dependence. This distribution could then be compared with other direct, dynamical 
determinations. Another situation worth investigating experimentally, suggested by the present study, 
is the out of equilibrium (aging) Kovacs effect, where both temperatures are kept well below 
the glass temperature $T_g$. 

Finally, from a theoretical point of view, it would be worth studying the predictions of the mean field (p-spin) spin-glass for the shape of the Kovacs hump. As is well known (see e.g.~\cite{Review}), the dynamical equations for this system are identical to the Mode-Coupling equations for structural glasses. Although we expect on general grounds that the results of this model should be again quite similar to those obtained in the present study, it would be interesting to check this assertion in more details.

\begin{table}
\begin{center}
\vspace{3mm}
\begin{tabular}{|c|c|c|c|}
\hline
& Domain growth ($\ell_{eq}=\infty$) & Trap model $T_1 > T_g$ & Trap model $T_1 < T_g$ \\ \hline
Preparation time $t_1$ & $\ell(t_1,T_1)^{\Theta-d} \propto T_2-T_1$ & $t_1^{1-\theta_1} \ln t_1 
\propto T_2-T_1$ & $t_w$ \\
\hline
Height of the hump $\Delta E_K$ & $T_2-T_1$ & $T_2-T_1$ & $T_2-T_1$ \\
\hline
Characteristic time $\tau^*$ & $\ell(\tau^*,T_2)=\ell(t_1,T_1)$ & $\tau^*=t_1^{T_1/T_2}$ 
& $\tau^*=t_w^{T_1/T_2}$ \\
\hline
Hump time $\tau_K$ & $\tau_K \sim (\tau^*)^{\frac{1}{\nu+1}} \ll \tau^*$ & $\tau_K \sim (\tau^*)^{\frac{1}{\nu+1}} \ll \tau^*$ 
& $\tau_K \sim \tau^*$ \\
\hline
$\Delta E$ at early time &$\Delta E_K\left(1-\ell^{\Theta-d}\right)$ & $\Delta E_K\left(1-t^{-\nu} \ln t \right)$
& $\Delta E_K (t/\tau^*)^{(1-\theta_1)/\gamma}$ \\
\hline
$\Delta E$ at late time & $\Delta E_K (\ell_1/\ell)^{d-\Theta}$ & $\Delta E_K (\tau^*/t)^\nu$ & $- T_2 \ln (t/t_w^{\gamma})$ \\
\hline
Exponent $\nu$ & $\nu=(d-\Theta)/z$ & $\nu = \theta_2-1$ & \\
\hline
\end{tabular}
\vspace{5mm}
\caption{Summary of the different results and regimes for the Kovacs hump, in the limit 
where $T_1 \to T_2^-$. We denote by $t_1$ or $t_w$ the time 
spent at the lowest temperature $T_1$, $\Theta$ and $z$ the energy and dynamical exponents for domain growth, $\gamma$ the ratio $\gamma=T_1/T_2$ and $\theta_{1,2}$ the reduced temperatures $\theta_{1,2}=T_{1,2}/T_g$.}
\label{tab-recap}
\end{center}
\end{table}

\section*{Acknowledgments} We want to thank E. Pitard and M. Sasaki 
(who both took part in an early stage of this work) and L. Berthier, L. Cugliandolo, J. Kurchan and 
E. Vincent for very useful discussions. We also thank a referee for pointing out a conceptual error in the Lennard-Jones simulation 
results presented in the first version of this manuscript.

\section*{Appendix: detailed calculation in the trap model}

\subsection{Probability distribution and Green function}

The Master equation of the trap model reads:

\be
\frac{\partial P_{\textsc{t}}}{\partial t}(E,t) = - e^{-\beta E} P_{\textsc{t}}(E) + \omega(t) \rho(E)
\ee
with $\omega(t)=\int_0^{\infty} dE' e^{-\beta E'} P_{\textsc{t}}(E',t)$, and $\beta=\frac{1}{T}$. $E$ is a positive variable, the depth of the traps, and is actually the opposite of the true energy of the states. This Master equation has to be supplemented by an initial condition:

\be
P_{\textsc{t}}(E,t=0) = P_0(E)
\ee
where $P_0(E)$ is a given (arbitrary) probability distribution. We also take an exponential density of states, $\rho(E)=T_g^{-1}\, e^{-E/T_g}$. Introducing the Laplace transform $\hat{P}_T(E,s)$ with respect to $t$ defined as:

\be
\hat{P}_{\textsc{t}}(E,s) = \int_0^{\infty} dt\, e^{-st} P_{\textsc{t}}(E,t)
\ee
the Master equation becomes:

\be
s \hat{P}_{\textsc{t}}(E,s) - P_0(E) = -e^{-\beta E} \hat{P}_{\textsc{t}}(E,s) + \hat{\omega}(s) \rho(E)
\ee
Solving for $\hat{P}_{\textsc{t}}(E,s)$, one has:

\be \label{eq31-PEs}
\hat{P}_{\textsc{t}}(E,s) = \frac{P_0(E)}{s+e^{-\beta E}} + \frac{\hat{\omega}(s) \rho(E)}{s+e^{-\beta E}}
\ee
$\hat{\omega}(s)$ is determined by multiplying Eq.~(\ref{eq31-PEs}) by $e^{-\beta E}$ and integrating over $E$. The distribution $\hat{P}_{\textsc{t}}(E,s)$ is then given by:

\be \label{eqn_PEs}
\hat{P}_{\textsc{t}}(E,s) = \frac{e^{\beta E}}{1+s\, e^{\beta E}}\, P_0(E) + \frac{1}{s}\, \frac{e^{\beta E} \rho(E)}{1+s\, e^{\beta E}}\, \hat{\varphi}(s) \left[\int_0^{\infty} dE\, \frac{e^{\beta E} \rho(E)}{1+s\, e^{\beta E}}\right]^{-1}
\ee
with $\hat{\varphi}(s)$ defined as:

\be
\hat{\varphi}(s)=\int_0^{\infty} dE \, \frac{P_0(E)}{1+s e^{\beta E}}
\ee
Integrating Eq.~(\ref{eqn_PEs}) over $E$ allows to check that $\hat{P}_{\textsc{t}}(E,s)$ is well normalized, i.e. $\int_0^{\infty} dE\, \hat{P}_{\textsc{t}}(E,s) = 1/s$. In order to compute the variation of the energy after a temperature shift, one has to introduce the Green function $G_{\textsc{t}}(E,E_0,t)$ defined as the probability for the system to have energy $E$ at time $t_w+t$ given that the energy was $E_0$ at time $t_w$, if the bath temperature is $T$. Note that since the process is Markovian, the Green function depends only on the time difference $t$, and not on $t_w$. The Green function in Laplace space $\hat{G}_{\textsc{t}}(E,E_0,s)$ is straightforwardly obtained from Eq.~(\ref{eqn_PEs}) choosing $P_0(E) = \delta(E-E_0)$:

\be
\hat{G}_{\textsc{t}}(E,E_0,s) = \frac{e^{\beta E_0}}{1+s\, e^{\beta E_0}}\, \delta(E-E_0) + \frac{1}{s}\, \frac{1}{1+s\,e^{\beta E_0}}\, \frac{e^{\beta E} \rho(E)}{1+s\, e^{\beta E}}\, \left[\int_0^{\infty} dE\, \frac{e^{\beta E} \rho(E)}{1+s\, e^{\beta E}}\right]^{-1}
\ee

As shown in \cite{Monthus}, the energy distribution $P_{\textsc{t}}(E,t)$ takes a scaling form for large times. Indeed, from Eq.~(\ref{eqn_PEs}), one has for $s \to 0$:

\be \label{eqn_PiEs}
\hat{P}_{\textsc{t}}(E,s) = \frac{1}{s}\, \frac{e^{\beta E} \rho(E)}{1+s\, e^{\beta E}}\, \left[\int_0^{\infty} dE\, \frac{e^{\beta E} \rho(E)}{1+s\, e^{\beta E}}\right]^{-1} = \frac{\sin \pi \theta}{\pi}\, \frac{\beta\, e^{\beta E}}{(1+s\, e^{\beta E})(s\, e^{\beta E})^{\theta}} \equiv \hat{\Pi}_{\textsc{t}}(E,s)
\ee
which defines the asymptotic distribution $\hat{\Pi}_{\textsc{t}}(E,s)$. The reduced temperature $\theta=T/T_g$ has also been introduced. Its inverse Laplace transform $\Pi_{\textsc{t}}(E,t)$ satisfies a scaling relation in the variable $\xi = \frac{e^{E/T}}{t}$:

\be
\Pi_{\textsc{t}}(E,t) = \beta\, \xi\, g(\xi)
\ee
One finds $g(\xi)$ by inverting the Laplace transform given by Eq.~(\ref{eqn_PiEs}):

\be
g(\xi) = \frac{\sin \pi \theta}{\pi\, \Gamma(\theta)}\, \frac{1}{\xi}\, e^{-1/\xi} \int_0^{1/\xi} du\, u^{\theta-1} e^u
\ee 
So for large times, the Green function is given by:

\be \label{eq38-GEE0s}
\hat{G}_{\textsc{t}}(E,E_0,s) = \frac{e^{\beta E_0}}{1+s\, e^{\beta E_0}}\, \delta(E-E_0) + \frac{1}{1+s\, e^{\beta E_0}}\, \hat{\Pi}_{\textsc{t}}(E,s)
\ee
We now consider the following thermal history: at the initial time, the system is quenched from $T_0>T_g$ to $T_1<T_g$; at time $t_w$, it is re-heated to a temperature $T_2$ satisfying $T_1<T_2<T_g$. One is interested in the subsequent evolution of the energy, at time $t_w+t$. The probability to have energy $E$ at time $t_w+t$, given this thermal history, is:

\be
P(E,t_w,t) = \int_0^{\infty} dE_0\, G_{\textsc{t}_2}(E,E_0,t)\, P_{\textsc{t}_1}(E_0,t_w)
\ee
Taking the double Laplace transform with respect to $t_w$ and $t$:
\be
\hat{P}(E,s_w,s) = \int_0^{\infty} dE_0\, \hat{G}_{\textsc{t}_2}(E,E_0,s)\, \hat{P}_{\textsc{t}_1}(E_0,s_w)
\ee
Using the asymptotic expressions Eqs.~(\ref{eqn_PiEs}) and (\ref{eq38-GEE0s}), one can write:

\be
\hat{P}(E,s_w,s) = \frac{e^{\beta_2 E}}{1+s\, e^{\beta_2 E}} \, \hat{\Pi}_{\textsc{t}_1}(E,s_w) + \hat{\Pi}_{\textsc{t}_2}(E,s) \int_0^{\infty} dE_0 \frac{\hat{\Pi}_{\textsc{t}_1}(E_0,s_w)}{1+s\,e^{\beta_2}E}
\ee

\subsection{Evolution of the average energy and scaling relation in the aging regime}

Bearing in mind that the energy of a given state is the opposite of the energy barrier $E$, the mean energy $\overline{E}(t)$ at time $t$ is defined by:

\be
-\overline{E}(t) = \int_0^{\infty} dE\, E\, P_{\textsc{t}}(E,t)
\ee
Taking into account the thermal history introduced in the preceding section, we define the energy variation between time $t_w$ and $t_w+t$:

\be
\Delta E(t_w,t) \equiv \overline{E}(t_w+t) - \overline{E}(t_w)
\ee
Computing the double Laplace transform yields:

\be
-\Delta \hat{E}(s_w,s) = \int_0^{\infty} dE \int_0^{\infty} dE_0 \, (E-E_0)\, \hat{G}_{\textsc{t}_2}(E,E_0,s)\, \hat{P}_{\textsc{t}_1}(E_0,s_w)
\ee
Expanding this equation, one finds for $s_w \tau_0$ and $s \tau_0 \ll 1$:

\bea \label{deltaE1}
-\Delta \hat{E}(s_w,s) &=& \int_0^{\infty} dE\,E\, \hat{\Pi}_{\textsc{t}_2}(E,s) \int_0^{\infty} dE_0\, \frac{\hat{\Pi}_{\textsc{t}_1}(E_0,s_w)}{1+s\,e^{\beta_2 E_0}} - \frac{1}{s} \int_0^{\infty} dE_0\, E_0\, \frac{\hat{\Pi}_{\textsc{t}_1}(E_0,s_w)}{1+s\,e^{\beta_2 E_0}}\\ \label{deltaE2}
&=& \hat{I}(s)\, \hat{J}(s_w,s)-\frac{1}{s} \hat{K}(s_w,s)
\eea
where $\hat{I}(s)$, $\hat{J}(s_w,s)$ and $\hat{K}(s_w,s)$ denote respectively the three integrals appearing in Eq.~(\ref{deltaE1}). Making the change of variable $\tau = e^{\beta_2 E}$ in $\hat{I}$ and $\tau=e^{\beta_1 E_0}$ in $\hat{J}$ and $\hat{K}$, one has:

\bea
\hat{I}(s) &=& \frac{\sin \pi \theta_2}{\pi} \int_1^{\infty} d\tau\, \frac{T_2 \ln \tau}{(1+s\tau)(s\tau)^{\theta_2}}\\
\hat{J}(s_w,s) &=& \frac{\sin \pi \theta_1}{\pi} \int_1^{\infty} \frac{d\tau}{(1+s\tau^{\gamma})(1+s_w\tau)(s_w\tau)^{\theta_1}}\\
\hat{K}(s_w,s) &=& \frac{\sin \pi \theta_1}{\pi} \int_1^{\infty} \frac{d\tau \, T_1 \, \ln \tau}{(1+s\tau^{\gamma})(1+s_w\tau)(s_w\tau)^{\theta_1}}
\eea
with $\gamma=T_1/T_2$; $\hat{I}(s)$ can be computed using the identity 
$\ln \tau = \partial \tau^{\alpha} / \partial \alpha_{\, | \alpha=0}$; one finds:

\be
\hat{I}(s)=\frac{T_2}{s} (\pi \cot \pi \theta_2 - \ln s)
\ee
Let us show that $\Delta \hat{E}(s_w,s)$ satisfies a scaling relation. Note first that for $s \to 0$, $\hat{J}(s_w,s)$ is of the form:

\be
\hat{J}(s_w,s) = \frac{1}{s^{1/\gamma}} \int_0^{\infty} du\, \frac{f(s_w u/s^{1/\gamma})}{1+u^{\gamma}}
\ee
where $f(x)=(\sin \pi \theta_1)/[\pi x^{\theta_1}(1+x)]$, and $u=s^{1/\gamma}\tau$.
In the same way, $\hat{K}(s_w,s)$ reads:

\be
\hat{K}(s_w,s)=-T_2 \ln s \, \hat{J}(s_w,s) + \frac{T_1}{s^{1/\gamma}} \int_0^{\infty} du \, \frac{f(s_w u/s^{1/\gamma})}{1+u^{\gamma}} \ln u
\ee
Coming back to $\Delta \hat{E}(s_w,s)$, one has from Eq.~(\ref{deltaE2}) that terms in $\ln s$ cancel, and one gets:

\be \label{deltaE3}
-\Delta \hat{E}(s_w,s) = \frac{1}{s^{1+1/\gamma}} \left[ \pi T_2 \cot \pi \theta_2 \int_0^{\infty} du\, \frac{f(s_w u/s^{1/\gamma})}{1+u^{\gamma}} - T_1 \int_0^{\infty} du\, \frac{f(s_w u/s^{1/\gamma})}{1+u^{\gamma}} \ln u \right] = \frac{1}{s^{1+1/\gamma}}\, \varphi \left(\frac{s_w}{s^{1/\gamma}} \right)
\ee
which implies a simple scaling form $\Delta E(t_w,t) = \psi(t/t_w^{\gamma})$. This is easily shown by computing the Laplace transform of this scaling form:

\be
{\cal L}_{t_w t} \psi \left(\frac{t}{t_w^{\gamma}}\right) = \int_0^{\infty} dt \int_0^{\infty} dt_w\, e^{-st}\, e^{-s_w t_w} \psi \left(\frac{t}{t_w^{\gamma}}\right)
\ee
Let us make the following changes of variable: $t=x t_w^{\gamma}$ (at fixed $t_w$), and then $t_w=v/(sx)^{1/\gamma}$ (at fixed $x$). One finally gets:

\be
{\cal L}_{t_w t} \psi \left(\frac{t}{t_w^{\gamma}}\right) = \frac{1}{s^{1+1/\gamma}} \int_0^{\infty} \frac{dx}{x^{1+1/\gamma}} \psi(x) \int_0^{\infty} dv\, v^{\gamma}\, \exp \left(-\frac{s_w}{s^{1/\gamma}} \frac{v}{x^{1/\gamma}} - v^{\gamma} \right) = \frac{1}{s^{1+1/\gamma}} \varphi \left(\frac{s_w}{s^{1/\gamma}} \right)
\ee
which indeed gives back the expected scaling form in Laplace space.

\subsection{Short time behaviour}

In this section, we shall focus on the short time behaviour of $\Delta E(t_w,t)$, characterized by $t \ll t_w^{\gamma}$, or equivalently $s \gg s_w^{\gamma}$. Note however that we consider only times that are large compared to the microscopic time scale: $t, t_w \gg \tau_0=1$ ($s, s_w \ll 1$). From Eq.~(\ref{deltaE3}), one sees that two integrals have to be computed:

\bea \label{Alambda}
A(\lambda) &=& \frac{\sin \pi \theta_1}{\pi} \int_0^{\infty} \frac{du}{(1+u^{\gamma})(1+\lambda u)(\lambda u)^{\theta_1}}\\ \label{Blambda}
B(\lambda) &=& \frac{\sin \pi \theta_1}{\pi} \int_0^{\infty} \frac{\ln u \,du}{(1+u^{\gamma})(1+\lambda u)(\lambda u)^{\theta_1}}
\eea
where $\lambda$ stands for the ratio $s_w/s^{1/\gamma}$. In the case $\lambda \ll 1$, these integrals reduce to:

\be
A(\lambda) = \frac{\sin \pi \theta_1}{\pi \lambda^{\theta_1}} \int_0^{\infty} \frac{du}{(1+u^{\gamma})u^{\theta_1}} \qquad B(\lambda) = \frac{\sin \pi \theta_1}{\pi \lambda^{\theta_1}} \int_0^{\infty} \frac{\ln u\, du}{(1+u^{\gamma})u^{\theta_1}}
\ee\
on condition that $\theta_1+\gamma > 1$. The opposite case, $\theta_1+\gamma < 1$, will be considered later on. The integrals $A(\lambda)$ and $B(\lambda)$ are readily calculated using the following identities:

\be
\int_0^{\infty} \frac{dv}{v^{\mu}(1+v)} = \frac{\pi}{\sin \pi \mu} \qquad 
\int_0^{\infty} \frac{\ln v\, dv}{v^{\mu}(1+v)} = \frac{\pi^2 \cos \pi \mu}{\sin^2 \pi \mu}
\ee
Altogether, one finds for $\Delta \hat{E}(s_w,s)$:

\be
-\Delta \hat{E}(s_w,s) \simeq \frac{T_2 \pi \sin \pi \theta_1}{\gamma \sin \frac{\pi}{\gamma}(1-\theta_1)} [\cot \frac{\pi}{\gamma}(1-\theta_1) + \cot \pi \theta_2] \frac{1}{s^{1+1/\gamma}} \left(\frac{s^{1/\gamma}}{s_w} \right)^{\theta_1}
\ee
The short time behaviour of $\Delta E(t_w,t)$ is obtained by inverse Laplace transform, in the case $\theta_1+\gamma > 1$:

\be \label{deltaE4}
\Delta E(t_w,t) \simeq K_> \, \left( \frac{t}{t_w^{\gamma}} \right)^{(1-\theta_1)/\gamma} \qquad t \ll t_w^{\gamma}
\ee
where the coefficient $K_>$ is given by:
\be \label{coef-K>}
K_> = - \frac{T_2 \pi \sin \pi \theta_1 [\cot \frac{\pi}{\gamma}(1-\theta_1) + \cot \pi \theta_2]}{\gamma \sin [\frac{\pi}{\gamma}(1-\theta_1)]\, \Gamma(\theta_1)\, \Gamma(\frac{1+\gamma-\theta_1}{\theta_1})}
\ee
Note that in spite of the minus sign in the r.h.s. of Eq.~(\ref{deltaE4}), $\Delta E(t_w,t)$ is indeed positive at short times for $\theta_2 > \theta_1$, showing that the energy has to increase first before reaching lower values. However, this coefficient vanishes for $\theta_2=\theta_1$ (i.e. no singularity occurs if temperature is kept constant), and becomes negative for $\theta_2 < \theta_1$. \par
\
In the opposite case, $\theta_1+\gamma < 1$, another approximation has to be used. Making the change of variable $v=\lambda u$ in $A(\lambda)$ and $B(\lambda)$ --see Eqs.~(\ref{Alambda},\ref{Blambda})-- one finds:

\bea
A(\lambda) &=& \frac{\sin \pi \theta_1}{\pi \lambda} \int_0^{\infty} \frac{dv}{[1+(v/\lambda)^{\gamma}](1+v) v^{\theta_1}}\\
B(\lambda) &=& \frac{\sin \pi \theta_1}{\pi \lambda} \int_0^{\infty} dv \frac{\ln v - \ln \lambda}{[1+(v/\lambda)^{\gamma}](1+v) v^{\theta_1}}
\eea
In the small $\lambda$ limit, $(v/\lambda)^{\gamma} \gg 1$, so that $A(\lambda)$ and $B(\lambda)$ reduce to:

\be
A(\lambda) \simeq \frac{\sin \pi \theta_1}{\pi \lambda^{1-\gamma}} \int_0^{\infty} \frac{dv}{v^{\gamma+\theta_1}(1+v)} \qquad
B(\lambda) \simeq \frac{\sin \pi \theta_1}{\pi \lambda^{1-\gamma}} \int_0^{\infty} dv \frac{\ln v -\ln \lambda}{v^{\gamma+\theta_1}(1+v)}
\ee
which are indeed convergent since $\gamma+\theta_1 < 1$. One then finds for $\Delta \hat{E}(s_w,s)$:

\be
-\Delta \hat{E}(s_w,s) \simeq \frac{\sin \pi \theta_1}{\sin \pi (\gamma+\theta_1)}\, \left[\pi T_2 \cot \pi \theta_2 - \pi T_1 \cot \pi (\gamma+\theta_1) + T_1 \ln \frac{s_w}{s^{1/\gamma}}\right] \frac{1}{s^{1+1/\gamma}} \left(\frac{s^{1/\gamma}}{s_w}\right)^{1-\gamma} 
\ee
The inverse Laplace transform yields:

\be
\Delta E(t_w,t) \simeq K_< \left(C-\ln \frac{t}{t_w^{\gamma}}\right)\, \frac{t}{t_w^{\gamma}} \qquad t \ll t_w^{\gamma}
\ee
where $K_<$ and $C$ are given by:

\bea \label{coef-K<}
K_< &=& \frac{T_2 \sin \pi\theta_1}{\Gamma(1-\gamma) \sin \pi (\gamma+\theta_1)} \\ \label{coef-C}
C &=& \gamma\frac{\Gamma'(1-\gamma)}{\Gamma(1-\gamma)} - \Gamma'(2) + \pi \cot \pi \theta_2 - \pi \cot \pi(\gamma+\theta_1)
\eea

\subsection{Long time behaviour}
One can also study the long time behaviour $t \gg t_w^{\gamma}$, which happens to be easier to handle than the short time one. Coming back to the starting equation~(\ref{deltaE1}), the limit $s \ll s_w^{\gamma}$ simplifies a lot the equation, and one gets:

\be
-\Delta \hat{E}(s_w,s) = \frac{T_2}{s} [\pi \cot \pi \theta_2 - \ln s] \underbrace{\int_0^{\infty} dE_0 \hat{\Pi}_{\textsc{t}_1}(E_0,s_w)}_{1/s_w} - \frac{1}{s} \int_0^{\infty} dE_0 \, E_0 \hat{\Pi}_{\textsc{t}_1}(E_0,s_w)
\ee
The second term is nothing but ${\cal L}_{t t_w} \overline{E}(t_w)$, which also appears in the left hand side of the equation, due to the definition of $\Delta \hat{E}(s_w,s)$. In other words, the long time behaviour of $\overline{E}(t_w+t)$ appears to be the same as if the system had been quenched from high temperature to $T_2$ at time $t_w$. One finally finds:

\bea
\Delta E(t_w,t) &=& \overline{E}(t_w+t) - \overline{E}(t_w)\\
\overline{E}(t_w+t) &=& -{\cal L}_t^{-1} \frac{T_2}{s} (\pi \cot \pi \theta_2 - \ln s)\\
&=& T_2\, [\Gamma'(1) - \pi \cot \pi \theta_2] - T_2 \ln t
\eea
showing that $\overline{E}(t_w+t)$ is indeed independent from $t_w$ and from $T_1$ for times $t \gg t_w^{\gamma}$. This result can also be found directly without using this particular thermal procedure: if one computes the probability distribution $P(E,t)$ for large times $t$, starting from an arbitrary initial distribution $P_0(E)$, it appears that the asymptotic (large $t$) distribution does not depend on $P_0(E)$:

\be
\overline{E}(t_w+t,t_w) = \overline{E}_{late}(t)
\ee
where $\overline{E}_{late}(t)$ is the average energy at a large time $t$ after a quench from high temperature.

\end{document}